# Analysis Co-Sparse Coding for Energy Disaggregation

Shikha Singh and Angshul Majumdar, *Senior Member IEEE*

***Abstract*— Energy disaggregation is the task of segregating the aggregate energy of the entire building (as logged by the smartmeter) into the energy consumed by individual appliances. This is a single channel (the only channel being the smart-meter) blind source (different electrical appliances) separation problem. In recent times dictionary learning based approaches have shown promise in addressing the disaggregation problem. The usual technique is to learn a dictionary for every device and use the learnt dictionaries as basis for blind source separation during disaggregation. Dictionary learning is a synthesis formulation; in this work, we propose an analysis approach. The advantage of our proposed approach is that, the requirement of training volume drastically reduces compared to state-of-the-art techniques. This means that, we require fewer instrumented homes, or fewer days of instrumentation per home; in either case this drastically reduces the sensing cost. Results on two benchmark datasets show that our method produces the same level of disaggregation accuracy as state-of-the-art methods but with only a fraction of the training data.**

***Index Terms*—Energy Disaggregation, Non-intrusive Load Monitoring, Dictionary Learning, Analysis.**

## I. Introduction

ENERGY disaggregation addresses the problem of breaking down the total load read by the meter into consumption of individual appliances. Since its onset, the topic has been treated as a classical machine learning problem. During the training phase appliance level data is collected to train a model (e.g. Factorial Hidden Markov Model or Synthesis Sparse Coding); when in operation (test phase) the learnt model is used to separate the aggregate data (as read by the meter) into individual appliance level components. During the operational stage, appliance level monitoring is not required any further; hence the term 'non intrusive load monitoring' (NILM).

The pioneering study on NILM [1] assumed there is marked difference in power consumption level of each appliance; hence the loads were modeled as finite state machines. More recent techniques account for noise in the system and employ stochastic finite state machines like Factorial Hidden Markov Model (FHMM) instead [2-4]. These techniques can disaggregate appliances that have marked change in power consumption among states, e.g. AC, Fan, CFL, Refrigerator, etc. but do not disaggregate continuously varying loads like computers, laptops and printers very well. However in spite of the shortcoming, HMM based techniques continue to be popular [5].

Sparse Coding based techniques [6, 7] are not dependent on the Markov jump assumption, and hence are theoretically more suitable for handling continuously time varying loads. A recent study, that combines deep learning and sparse coding [8] is known to yield the best results when copious volume of training data is available.

Earlier studies in NILM can be broadly divided into two approaches – steady state analysis and transient state analysis. The works of Sultanem [9], Marceau and Zmeureanu [10] and Laughman et al [11] apply steady state analysis for energy disaggregation. These studies were based on edge detection (steady state power changes). Other techniques based on steady state used harmonic analysis, e.g., Nakano [12] and Leeb at al [13] belongs to this category; use of harmonics added extra information to the signature.

Transient state analysis refers to the branch of work that studies the noise in the system introduced by the change in the appliance's state. Studies like [14-16] belong to this category. The main disadvantage of these methods is that they require high frequency data. This is not practical in most situations. However alternative (to smart-meters) instrumentation techniques is an active area of research [17-19].

Another class of techniques that is gaining popularity is the multi-label classification approach [20-22]. In these methods, one cannot directly predict the energy consumption of individual appliances but figure out the appliances that are ON.

Recent studies in this area address scenarios that are significantly more challenging than traditional NILM. For example [23] proposes a training free solution using state-of-the-art graph signal processing techniques. Other studies attempt estimating the consumption of each appliance only from the monthly power bill [24, 25]. However, these studies are not directly relevant to our work, and hence will not be discussed in detail. We restrict our work to the more traditional NILM scenario.

There are many other techniques that have been proposed to solve NILM over the years; e.g. neural network based methods [26, 27], genetic algorithm [28], fuzzy logic [29] etc. These techniques are not directly relevant to our work. The interested reader can peruse the most recent survey on this topic [30].

Our work is based out of the dictionary learning / sparse coding approach. In the past a synthesis dictionary learning based method has been used for modelling the devices. In this work we propose an analysis approach. Our work follows from the success of analysis dictionary learning [31] over its synthesis counterpart in image processing. The main advantage of the analysis formulation is that, it is less prone to over-fitting. This would mean that we need less training data from households. In other words, it means that our method would require installing sensors on fewer homes, or installing them for fewer days on the same home. In either case, this brings down the cost of sensor deployment drastically.

We give a hypothetical example to elucidate our work's impact. Consider energy disaggregation of 10 households on a block. Say there are 10 appliances in the house, therefore the cost of instrumenting them will be around 150 USD per appliance and a Raspberry Pi controller per house (costing 100 USD); therefore the total cost of instrumenting per house would be around 1600 USD. For disaggregating 10 houses, existing techniques (as shown in the experimental section) will need to instrument at least 5, therefore the total cost of instrumentation would be 8000 USD. With our method, we would only need instrumenting a single house, therefore the cost reduces by 5-fold. This cost reduction is only for a small block; at a larger scale the cost saving would be even more pronounced.

The rest of the paper is organised into several sections. Literature on synthesis and analysis dictionary learning is reviewed in the following section. The proposed technique is described in section 3. The results on benchmark datasets are reported in section 4. The conclusions of this work are discussed in section 5.

## II. Literature Review

### A. Synthesis Sparse Coding

Dictionary learning is a synthesis approach. Given the data (*X*) it learns a single level dictionary (*D*) such that the it can regenerate / synthesize the data from the learnt coefficients (*Z*). This is expressed as,

$$X = DZ \quad (1)$$

This (1) is a typical matrix factorization problem. In sparse coding, the objective is to have a sparse coefficients matrix Z. K-SVD [32] is perhaps the most popular algorithm for sparse coding. It solves a problem of the form –

$$\min_{D,Z} \|X - DZ\|_F^2 \text{ s.t.} \|Z\|_0 \le \tau \quad (2)$$

We have abused the notation slightly; the $l_0$-norm is defined on the vectorised version of the coefficient *Z*. Here $\tau$ is defines the sparsity level of the coefficients.

K-SVD is a good algorithm, but is relatively slow owing to the necessity of computing SVDs in every iteration and running a slow orthogonal matching pursuit algorithm for sparse coding. Practical applications of dictionary learning solve an unconstrained version of (2) with an $l_1$-norm for promoting sparsity.

Dictionary learning has enjoyed immense popularity in the last decade. It has been the de facto tool for solving many inverse problems in signal and image processing. Machine learning researchers used supervised variants of dictionary learning for many computer vision problems.

The application of dictionary learning in NILM was introduced by Kolter et al [6]. It makes the standard assumption that there is training data collected over time, where the smartmeter logs only consumption from a single device only. This can be expressed as $X_i$ where *i* is the index for an appliance, the columns of $X_i$ are the readings over a period of time. For each appliance they learnt a dictionary, i.e. they expressed:

$$X_i = D_i Z_i,\ i = 1...\text{N} \quad (3)$$

where $D_i$ represents the basis/dictionary, $Z_i$ are the loading coefficients, assumed to be sparse and *N* is the total number of appliances.

Instead of using K-SVD (for reasons mentioned before), [6] formulated learning in the following fashion.

$$\min_{D_i, Z_i} \|X_i - D_i Z_i\|_F^2 + \lambda \|Z_i\|_1,\ i = 1...\text{N} \quad (4)$$

Positivity constraint is enforced on the coefficients. Note that the $l_1$-norm replaces the prior (2) $l_0$-norm, since the former is more efficient to implement.

This is a basic sparse coding formulation. In [6] some discriminative penalties were enforced on (4) to slightly improve the results. Extensions of this work [7] impose dynamical model (like autoregression) on the dictionary atoms. Other such minute variations exists as well.

During actual operation, several appliances are likely to be in use simultaneously. In such a case (assuming passive loads) the aggregate power read by the smartmeter is a sum of the powers for individual appliances. Thus if *X* is the total power from N appliances (where the columns indicate smartmeter readings over the same period of time as in training) the aggregate power can be modeled as:

$$X = \sum_i X_i = \sum_i D_i Z_i \quad (5)$$

Here the summation is over all the N appliances.

Given this model, it is possible to find out the loading coefficients of each device by solving the following sparse recovery problem,

$$\min_{Z_1,...,Z_N} \left\| X - \left[ D_1 | ... | D_N \right] \begin{bmatrix} Z_1 \\ ... \\ Z_N \end{bmatrix} \right\|_F^2 + \lambda \left\| \begin{matrix} Z_1 \\ ... \\ Z_N \end{matrix} \right\|_1 \quad (6)$$

Here a positivity constraint on the loading coefficients is enforced as well. This is a convex problem since the basis are fixed. Once the loading coefficients are estimated, one can easily compute the power consumption from individual devices.

$$\hat{X}_i = D_i Z_i,\ i = 1...\text{N} \quad (7)$$

A deep extension to sparse coding has been proposed in [8]. Instead of learning a single level of dictionary, multiple levels of dictionaries are being learnt for each appliance, i.e. instead of (3), it proposes learning three levels; this is expressed as,

$$X_i = D_i^1 D_i^2 D_i^3 Z_i,\ i = 1...\text{N} \quad (8)$$

During the learning stage, multiple levels of dictionaries needed to be solved. In [8], this is expressed as,

$$\min_{D_i^1, D_i^2, D_i^3, Z_i} \|X_i - D_i^1 D_i^2 D_i^3 Z_i\|_F^2 + \lambda \|Z_i\|_1,\ i = 1...\text{N} \quad (9)$$

A Bregman Splitting techniques was used in [8] to solve (9).

Once the multi-level dictionaries are learnt, the disaggregation phase remains the same as before.

$$X = \sum_i X_i = \sum_i D_i^1 D_i^2 D_i^3 Z_i \tag{10}$$

Once learnt, the dictionaries can be collapsed into a single level to have exactly the same form as (5). Therefore the loading coefficients during the test stage can be obtained by solving (6). Once the coefficients are obtained, the energy consumption can be found using (7).

### B. Analysis Co-Sparse Coding

In co-sparse analysis dictionary learning [31], the signal is analysed to generate the sparse coefficients. The solution is framed such that the sparse coefficients are not obtained; rather a clean version of the data ( $\hat{X}$ ) is obtained so that, when operated on by the analysis dictionary (*D*), sparse coefficients are produced. Mathematically the learning is represented as,

$$\min_{D,\hat{X}} \left\| X - \hat{X} \right\|_F^2 \text{ s.t.} \left\| D\hat{X} \right\|_0 \le \tau \tag{11}$$

Here *D* is the analysis dictionary; it is different from the synthesis dictionary of (1). There should not be any confusion between the two, since the context / model is different.

The analysis K-SVD algorithm is not as popular as its synthesis counterpart is mainly because it has an inefficient implementation. But it enjoys nice theoretical advantage over its synthesis counterpart.

A little analysis shows that for a synthesis dictionary of size $m \times n$, with sparsity (number of non-zero elements in *Z*) *k*, the number of sub-spaces is ${}^nC_k$ for k-dimensional sub-spaces. For analysis dictionary learning of size $p \times d$, with co-sparsity *l* the number of sub-spaces is ${}^pC_l$ for sub-spaces of dimension *d-l*. If we assume equal redundancy, i.e. *p=n=2d*, and equal dimensionality of the sub-space, i.e. *k=d-l*, the number of analysis sub-spaces will be *n* where as the number of synthesis sub-spaces are $k\log_2(n/k)$ (via Stirling's approximation); usually $n \gg$ $n/k)$. For example with *n=700, l=300 and k = 50*, the number analysis sub-spaces are 700 whereas the number of synthesis sub-spaces are only 191.

This analysis means that for an analysis and a synthesis dictionary of same dimensions, an analysis dictionary is able to capture significantly more variability in the data compared to its synthesis counterpart. In other words, for a fixed training set a smaller sized transform need to be learned compared to a dictionary. From the machine learning perspective, given the limited training data, learning fewer parameters for the transform has less chance of over-fitting than learning a larger number of synthesis dictionary atoms. Hence, for limited training data, as is the case with most practical problems, transform learning can be assumed to yield better generalizability (and hence better results) compared to dictionary learning. This is the motivation behind our analysis formulation.

Analysis dictionary learning has only seen a handful of applications in the past [33, 34]. But wherever they have been used (super-resolution [33], MRI reconstruction [34]), they have surpassed synthesis dictionary learning formulations. Success of such practical studies also motivates this work.

## III. Proposed Analysis Formulations

The basic problem statement remains the same as in the synthesis case. There is training data available for each device ($X_i$); along the rows it denotes the time period and along the columns it denotes the days. In this work, we modify from the synthesis to the analysis formulation. We propose three algorithms for analysis sparse coding with increasing levels of complexity. We start with the basic formulation.

### A. Simple Co-Sparse Coding

Just as in sparse synthesis coding, for each appliance (*i*), we learn an analysis dictionary –

$$\min_{D_i,\hat{X}_i} \left\| X_i - \hat{X}_i \right\|_F^2 + \lambda \left\| D_i \hat{X}_i \right\|_1 \tag{12}$$

Note that we have using an unconstrained formulation; and employing $l_1$-norm for sparsity in place of the $l_0$-norm. The changes have been made to solve the problem more efficiently.

We propose a variable splitting technique to solve (12). We introduce a proxy $Z_i = D_i \hat{X}_i$. With this substitution, (12) is expressed as,

$$\min_{D_i,\hat{X}_i,Z_i} \left\| X_i - \hat{X}_i \right\|_F^2 + \lambda \left\| Z_i \right\|_1 \text{ s.t. } Z_i = D_i \hat{X}_i \tag{13}$$

Solving the exact Lagrangian for (13) is not necessary. It enforces exact equality between the variable and its proxy in each iteration. This is not required; we only want exact equality during convergence. Therefore, we formulate the augmented Lagrangian instead.

$$\min_{D_i,\hat{X}_i,Z_i} \left\| X_i - \hat{X}_i \right\|_F^2 + \lambda \left\| Z_i \right\|_1 + \mu \left\| Z_i - D_i \hat{X}_i \right\|_F^2 \tag{14}$$

The hyper-parameter $\mu$ controls the degree of equality between the variable and the proxy. For a small value, the constraint is relaxed and for a high value, equality is enforced. Usually a heuristic 'heating' technique is followed where one starts with a small value of $\mu$ and progressive increases it after solving (14).

The Split Bregman technique is a better alternative to such heuristic hyper-parameter heating. It has been used profusely in signal processing literature in the recent past (e.g. [8, 34]). This technique introduces a Bregman relaxation variable ($B_i$) in the constraint, leading to the following,

$$\min_{D_i,\hat{X}_i,Z_i} \left\| X_i - \hat{X}_i \right\|_F^2 + \lambda \left\| Z_i \right\|_1 + \mu \left\| Z_i - D_i \hat{X}_i - B_i \right\|_F^2 \tag{15}$$

Here the relaxation variable is updated in every iteration. Therefore there is no need to tune the hyper-parameter progressively. The Bregman variable automatically updates itself to enforce convergence between the variable and its proxy; one only needs to fix the hyper-parameter $\mu$ at a moderate value without much tuning.

The formulation (15) can be solved using the alternating direction method of multipliers [35, 36]. It can be segregated into the following sub-problems –

$$\text{P1:} \min_{D_i} \left\| Z_i - D_i \hat{X}_i - B_i \right\|_F^2$$

$$\text{P2:}\min_{\hat{X}_i} \left\| X_i - \hat{X}_i \right\|_F^2 + \mu \left\| Z_i - D_i \hat{X}_i - B_i \right\|_F^2$$

$$\equiv \min_{\hat{X}_i} \left\| \begin{pmatrix} X_i \\ \sqrt{\mu}\left(Z_i - B_i\right) \end{pmatrix} - \begin{pmatrix} I \\ \sqrt{\mu} D_i \end{pmatrix} \hat{X}_i \right\|_F^2$$

$$\text{P3:}\min_{Z_i} \left\| Z_i - D_i \hat{X}_i - B_i \right\|_F^2 + \frac{\lambda}{\mu} \left\| Z_i \right\|_1$$

Sub-problems P1 and P2 are simple least squares problems. They can be solved using closed form (Moore Penrose Pseudoinverse). Sub-problem P3 also has a closed form solution via soft thresholding.

$$Z_i \leftarrow signum\left(D_i \hat{X}_i + B_i\right) \cdot \max\left(\left|D_i \hat{X}_i + B_i\right| - \frac{\lambda}{2\mu}, 0\right) \quad (16)$$

The final step is to update the Bregman relaxation variable.

$$B_i \leftarrow Z_i - D_i \hat{X}_i - B_i \quad (17)$$

This concludes the steps per iteration. We can see that all the steps have efficient closed form solutions. This makes our solution significantly less time consuming (by several orders of magnitude) compared to the A-KSVD algorithm proposed in [31]. Our entire algorithm is succinctly given below.

For every appliance $i$ solve: $\min_{D_i, \hat{X}_i} \left\| X_i - \hat{X}_i \right\|_F^2 + \lambda \left\| D_i \hat{X}_i \right\|_1$

Initialize: $\hat{X}_i (= X_i)$ , $B_i = \mathbf{1}$ and $D_i$ randomly

Until convergence solve following three sub-problems in every loop

$$\text{P1:}\min_{D_i} \left\| Z_i - D_i \hat{X}_i - B_i \right\|_F^2$$

$$\text{P2:}\min_{\hat{X}_i} \left\| \begin{pmatrix} X_i \\ \sqrt{\mu}\left(Z_i - B_i\right) \end{pmatrix} - \begin{pmatrix} I \\ \sqrt{\mu} D_i \end{pmatrix} \hat{X}_i \right\|_F^2$$

$$\text{P3:}\min_{Z_i} \left\| Z_i - D_i \hat{X}_i - B_i \right\|_F^2 + \frac{\lambda}{\mu} \left\| Z_i \right\|_1$$

For disaggregation, we follow the standard model, i.e. the total power is the sum of the individual powers.

$$X = \sum_i X_i$$

As before, the summation is over the N appliances.

The goal is to recover the individual components ($X_i$'s) given the learnt analysis dictionaries.

We formulate disaggregation as,

$$\min_{\hat{X}_i\text{'s}} \left\| X - \sum_i \hat{X}_i \right\|_F^2 + \lambda \sum_i \left\| D_i \hat{X}_i \right\|_1 \quad (18)$$

Unlike the training phase, this (18) is a convex formulation.

Using alternating minimization (for each component), iteration '$k$' can be expressed as,

$$\min_{\hat{X}_i} \left\| X - \sum_{j \neq i} \hat{X}_j^{(k)} - \hat{X}_i \right\|_F^2 + \lambda \left\| D_i \hat{X}_i \right\|_1 \quad (19)$$

Here $\hat{X}_j^{(k)}$ denotes the component corresponding to the $j^{th}$ appliance that is not being updated in this sub-problem; they are all constants for this sub-problem.

The said problem (18) is a typical total variation type minimization problem [36]. However, such majorization minimization based techniques are inefficient. Today most studies employ the Split Bregman technique (e.g. [37]) for solving such problems. We follow the same in this work.

As in the training phase, we substitute $Z_i = D_i \hat{X}_i$ . After introducing the Bregman relaxation variable in the approximate equality constraint of the augmented Lagrangian formulation, we arrive at the following formulation,

$$\min_{\hat{X}_i, Z_i} \left\| X - \sum_{j \neq i} \hat{X}_j^{(k)} - \hat{X}_i \right\|_F^2 + \lambda \left\| Z_i \right\|_1 + \mu \left\| Z_i - D_i \hat{X}_i - B_i \right\|_F^2 \quad (20)$$

Using alternating minimization, the sub-problems are:

$$\text{P1:}\min_{\hat{X}_i} \left\| X - \sum_{j \neq i} \hat{X}_j^{(k)} - \hat{X}_i \right\|_F^2 + \mu \left\| Z_i - D_i \hat{X}_i - B_i \right\|_F^2$$

$$\equiv \min_{\hat{X}_i} \left\| \begin{pmatrix} X - \sum_{j \neq i} \hat{X}_j^{(k)} \\ \sqrt{\mu}\left(Z_i - B_i\right) \end{pmatrix} - \begin{pmatrix} I \\ \sqrt{\mu} D_i \end{pmatrix} \hat{X}_i \right\|_F^2$$

$$\text{P2:}\min_{Z_i} \left\| Z_i - D_i \hat{X}_i - B_i \right\|_F^2 + \frac{\lambda}{\mu} \left\| Z_i \right\|_1$$

We have already discussed the solution of these two sub-problems in the training phase. Both have closed form solutions. As before, the final step is to update the Bregman relaxation variables.

Note that in the testing phase, (19) is for solving the power consumption from only a single appliance. The same need to be repeated for every appliance within one loop. The complete algorithm for disaggregation is given below.

Initialize: $X_i^{(0)}$'s

Until convergence

In iteration 'k'

For every $i$ solve: $\min_{\hat{X}_i} \left\| X - \sum_{j \neq i} \hat{X}_j^{(k)} - \hat{X}_i \right\|_F^2 + \lambda \left\| D_i \hat{X}_i \right\|_1$

$$\text{P1:}\min_{\hat{X}_i} \left\| \begin{pmatrix} X - \sum_{j \neq i} \hat{X}_j^{(k)} \\ \sqrt{\mu}\left(Z_i - B_i\right) \end{pmatrix} - \begin{pmatrix} I \\ \sqrt{\mu} D_i \end{pmatrix} \hat{X}_i \right\|_F^2$$

$$\text{P2:}\min_{Z_i} \left\| Z_i - D_i \hat{X}_i - B_i \right\|_F^2 + \frac{\lambda}{\mu} \left\| Z_i \right\|_1$$

$$B_i \leftarrow Z_i - D_i \hat{X}_i - B_i$$

End iteration 'k'

### *B. Distinctive Dictionaries*

In our second formulation, we want to make the analysis dictionaries distinctive from each other, i.e. the dictionaries for each appliance should look different from others. This has not been made explicit in our first formulation.

To achieve this, we draw from literature on incoherent dictionary learning [39-41]. Note that for two similar dictionaries ($D_i$ and $D_j$) the inner product between the one and the other $D_i^T D_j$ will have high values along the diagonals and low values in the off diagonal elements. This property has been used in [39-41] to pose $\|D^T D - I\|_F^2$ as the incoherence penalty. In this work, we want to minimize the similarities between dictionaries of different appliances. Therefore, we impose a penalty of the form $\|D_i^T D_j - I\|_F^2$. Adding this penalty to the training phase leads to,

$$\min_{D_i\text{'s},\hat{X}_i\text{'s}} \sum_i \|X_i - \hat{X}_i\|_F^2 + \lambda \|D_i \hat{X}_i\|_1 + \eta \sum_{j \neq i} \|D_i^T D_j - I\|_F^2 \quad (20)$$

Notice that, unlike the previous formulation, where the appliance wise dictionaries were solved separately, the present formulation is coupled and hence all of them need to be solved simultaneously. (2) can also be expressed as,

$$\min_{D_i\text{'s},\hat{X}_i\text{'s}} \sum_i \|X_i - \hat{X}_i\|_F^2 + \lambda \|D_i \hat{X}_i\|_1 + \eta \|D_i^T D_{i^C} - I_{N-1}\|_F^2 \quad (21)$$

Here $D_{i^C}$ consists of all the dictionaries except the ith one stacked vertically one after the other; $I_{N-1}$ is simply the identity matrix repeated *N-1* times (where N is the number of appliances).

As before, using the same substitutions ( $Z_i = D_i \hat{X}$ ) of (15) and introducing the relaxation variable, we recast (21) in the Split Bregman formulation.

$$\min_{D_i\text{'s},\hat{X}_i\text{'s},Z_i\text{'s}} \sum_i \begin{matrix} \|X_i - \hat{X}_i\|_F^2 + \lambda \|Z_i\|_1 + \eta \|D_i^T D_{i^C} - I_{N-1}\|_F^2 \\ + \mu \|Z_i - D_i \hat{X}_i - B_i\|_F^2 \end{matrix} \quad (22)$$

The updates for $\hat{X}_i$ and $Z_i$ can be decoupled and hence remain the same as in the solution for (15); both of them are known to have closed form updates. The change is in the solution for the dictionaries. For each appliance, one needs solving,

$$\min_{D_i} \|Z_i - D_i \hat{X}_i - B_i\|_F^2 + \eta \|D_i^T D_{i^C} - I_{N-1}\|_F^2 \quad (23)$$

This is a simple least squares problem having a closed form solution.

The final step is to update the Bregman relaxation variable; it remains the same as before. Succinctly, the training algorithm is expressed as,

For all appliances solve:

$$\min_{D_i\text{'s},\hat{X}_i\text{'s},Z_i\text{'s}} \sum_i \begin{matrix} \|X_i - \hat{X}_i\|_F^2 + \lambda \|Z_i\|_1 + \eta \|D_i^T D_{i^C} - I_{N-1}\|_F^2 \\ + \mu \|Z_i - D_i \hat{X}_i - B_i\|_F^2 \end{matrix}$$

Initialize: $\hat{X}_i (= X_i)$ , $B_i$ = **1** and $D_i$ randomly

Until convergence solve following three sub-problems in every loop for each *i*

$$\min_{D_i} \|Z_i - D_i \hat{X}_i - B_i\|_F^2 + \eta \|D_i^T D_{i^C} - I_{N-1}\|_F^2$$

$$\min_{\hat{X}_i} \left\| \begin{pmatrix} X_i \\ \sqrt{\mu}(Z_i - B_i) \end{pmatrix} - \begin{pmatrix} I \\ \sqrt{\mu} D_i \end{pmatrix} \hat{X}_i \right\|_F^2$$

$$\min_{Z_i} \|Z_i - D_i \hat{X}_i - B_i\|_F^2 + \frac{\lambda}{\mu} \|Z_i\|_1$$

Once the dictionaries are learnt during the training phase, there is no change in the disaggregation stage. It remains exactly the same as before.

### *C. Disaggregating Dictionaries*

For disaggregation, we want the dictionaries corresponding to one particular appliance to express the data in a co-sparse fashion; the same dictionary should not sparsely represent data signals from other appliances. Therefore, we impose an $l_1$-norm on $D_i \hat{X}_i$ but a non-sparse (dense) $l_2$-norm on $D_j \hat{X}_i$ . We impose these penalties on top of the distinctive penalty we introduced in the last sub-section. Our formulation becomes,

$$\min_{D_i\text{'s},\hat{X}_i\text{'s}} \sum_i \begin{matrix} \|X_i - \hat{X}_i\|_F^2 + \lambda \|D_i \hat{X}_i\|_1 + \eta \|D_i^T D_{i^C} - I_{N-1}\|_F^2 \\ + \gamma \|D_{i^C}^T \hat{X}_i\|_F^2 \end{matrix} \quad (24)$$

The notation $D_{i^C}$ has already been defined before. The term $\|D_{i^C}^T \hat{X}_i\|_F^2$ promotes dense coefficients when dictionaries corresponding to other appliances are used.

Using the same proxy $Z_i = D_i \hat{X}$ , and relaxing the equality constraint between the variable and proxy in by a relaxation variable, we arrive at the following,

$$\min_{D_i\text{'s},Z_i\text{'s},\hat{X}_i\text{'s}} \sum_i \begin{matrix} \|X_i - \hat{X}_i\|_F^2 + \lambda \|Z_i\|_1 + \eta \|D_i^T D_{i^C} - I_{N-1}\|_F^2 \\ + \gamma \|D_{i^C}^T \hat{X}_i\|_F^2 + \mu \|Z_i - D_i \hat{X}_i - B_i\|_F^2 \end{matrix} \quad (25)$$

Using ADMM, (25) can be split into the following sub-problems.

$$\text{P1:} \min_{\hat{X}_i} \|X_i - \hat{X}_i\|_F^2 + \gamma \|D_{i^C}^T \hat{X}_i\|_F^2 + \mu \|Z_i - D_i \hat{X}_i - B_i\|_F^2$$

$$\text{P2:} \min_{Z_i} \|Z_i - D_i \hat{X}_i - B_i\|_F^2 + \frac{\lambda}{\mu} \|Z_i\|_1$$

$$\text{P3:} \min_{D_i\text{'s}} \sum_i \eta \|D_i^T D_{i^C} - I_{N-1}\|_F^2 + \mu \|Z_i - D_i \hat{X}_i - B_i\|_F^2$$

$$\equiv \min_{D_i} \eta \|D_i^T D_{i^C} - I_{N-1}\|_F^2 + \mu \|Z_i - D_i \hat{X}_i - B_i\|_F^2 \ \forall i$$

Since for updating $\hat{X}_i$ , the $D_i$'s are assumed to be constant, hence we are able to decouple P1 into individual appliances. The update for the proxy $Z_i$'s are always decoupled (P2). For updating $D_i$, all the other dictionaries are assumed to be fixed and hence we can decouple P3 to its equivalent form.

Sub-problem P1 is a simple least squares problem. It can be expressed as follows,

$$\min_{\hat{X}_i} \left\| \begin{pmatrix} X_i \\ 0 \\ \sqrt{\mu}(Z_i - B_i) \end{pmatrix} - \begin{pmatrix} I \\ \sqrt{\gamma} D_{i^C}^T \\ \sqrt{\mu} D_i \end{pmatrix} \hat{X}_i \right\|_F^2 \quad (26)$$

It has a closed form solution in the form of Moore-Penrose pseudoinverse.

The solution of sub-problem P2 has already been discussed in the first sub-section. IT requires only one step of soft thresholding. Sub-problem P3 remain the same as (23); it is a least squares problem having a closed form solution.

The last step in every iteration is to update the Bregman relaxation variable. We have already discussed that. This concludes the training stage. The algorithm is shown succinctly in the following box.

For all appliances solve:

$$\min_{D_i\text{'s}, Z_i\text{'s}, \hat{X}_i\text{'s}} \sum_i \begin{matrix} \left\| X_i - \hat{X}_i \right\|_F^2 + \lambda \left\| Z_i \right\|_1 + \eta \left\| D_i^T D_{i^C} - I_{N-1} \right\|_F^2 \\ + \gamma \left\| D_{i^C}^T \hat{X}_i \right\|_F^2 + \mu \left\| Z_i - D_i \hat{X}_i - B_i \right\|_F^2 \end{matrix}$$

Initialize: $\hat{X}_i (= X_i)$, $B_i = \mathbf{1}$ and $D_i$ randomly

Until convergence solve following three sub-problems in every loop for each $i$

$$\min_{D_i} \left\| Z_i - D_i \hat{X}_i - B_i \right\|_F^2 + \eta \left\| D_i^T D_{i^C} - I_{N-1} \right\|_F^2$$

$$\min_{\hat{X}_i} \left\| \begin{pmatrix} X_i \\ 0 \\ \sqrt{\mu}(Z_i - B_i) \end{pmatrix} - \begin{pmatrix} I \\ \sqrt{\gamma} D_{i^C}^T \\ \sqrt{\mu} D_i \end{pmatrix} \hat{X}_i \right\|_F^2$$

$$\min_{Z_i} \left\| Z_i - D_i \hat{X}_i - B_i \right\|_F^2 + \frac{\lambda}{\mu} \left\| Z_i \right\|_1$$

For disaggregation, there is no change from the first formulation given in section 3.1.

## IV. Results

In this work, we evaluate our proposed algorithm on benchmark datasets. Since it is a new method, carrying out experiments on open source benchmarks is more reproducible. In this work we evaluate on two popular datasets – REDD and Pecan Street.

### *A. REDD Dataset*

The REDD dataset [42] – a moderate size publicly available dataset for electricity disaggregation. The dataset consists of power consumption signals from six different houses, where for each house, the whole electricity consumption as well as electricity consumptions of about twenty different devices are recorded. The signals from each house are collected over a period of two weeks with a high frequency sampling rate of 15kHz. To prepare training and testing data, aggregated and sub-metered data are averaged over a time period of 10 minutes. In the standard evaluation protocol, the 5th house is omitted since the data from this one is insufficient.

The disaggregation accuracy is defined by [42] as follows,

$$Acc = 1 - \frac{\sum_t \sum_n \left| \hat{y}_t^{(i)} - y_t^{(i)} \right|}{2 \sum_t \bar{y}_t}$$

where $t$ denotes time instant and $n$ denotes a device; the '2' in the denominator is to discount the fact that the absolute value will "double count" errors.

We compare the performance of our proposed method with two baseline techniques - Factorial HMM (FHMM) [2], and discriminating sparse coding (discSC) [6]. The rest are state-of-the-art methods – Power Disaggregagtion (PED) [7], multi-label classification (MLC) [22], and Deep Sparse Coding (DSC) [8]. FHMM and discSC are standardized techniques and the parameters are known from the non intrusive load monitoring toolkit[1]. For the remaining, the parameter values have been obtained from the corresponding studies.

For our proposed technique, the value of the sparsity inducing parameter $\lambda$ has always been kept at 0.1 (for all algorithms). For the incoherence term, the parameter $\eta$ has been set to 0.2 and for the disaggregating term the value of $\gamma$ has been fixed at 0.05. These values were obtained by cross validation on the training data using the greedy L-curve technique. It is greedy in the sense, that the value of the common sparsity parameter is obtained from the first technique. It is kept fixed in the second technique to find out η; the values of $\lambda$ and $\eta$ have been fixed for the third formulation for fixing $\gamma$. Our algorithm is not sensitive to the value of the hyper-parameter for a wide range of values (between 0.01 and 0.95) – this is expected since the Bregman relaxation variable adjusts automatically. The number of atoms we have used for each device is 3.

As outlined by [42] – there are two protocols for evaluation. In the first one (called 'training), a portion of the data from every household is used as training samples and rest (from those households) is used for prediction. Usually 80% of the data (sequentially) is used for training and the remaining for testing. In the second mode, the data from four households are used for training and the remaining one is used for prediction (called 'testing'. The usual protocol for the testing mode is to use 4 houses for training the and 5th house for testing.

In this work we have argued that the motivation for using analysis dictionary learning is its generalization ability, one requires lesser training data. Therefore, we propose more challenging protocols for testing and training modes. For the testing mode, we will use only one of the houses for training and the remaining four for testing. In the training mode we use 20% of the data for (for each house) training and the remainder for testing. The split into training and testing set has been done randomly and 100 such splits have been made. We report the mean from all the splits. In the following table (I) we show results for training mode. The testing mode disaggregation accuracy is shown in Table II. For both the tables 'Simple' is the technique proposed in section III.A;

[1] https://github.com/nilmtk/nilmtk

'Distinctive' is the technique proposed in section III.B; and 'Disaggregating' is the technique proposed in section III.C.

The results conclusively show that our proposed methods are significantly better than others. Our baseline 'Simple' technique yields more than 5% improvement over the next best (PED) for the training mode and deep sparse coding (DSC) for the testing. What is interesting to note is that, a state-of-the-art deep learning algorithm like DSC performs the worst. This is because, deep learning is data hungry; in limited data settings such as the training, it overfits and performs significantly worse compared to other shallow techniques. However in the testing mode, since the data from all houses are aggregated, it performs better.

TABLE I
TRAINING MODE DISAGGREGATION ACCURACY (MEAN OF 4 TEST HOUSES)

| House | FHMM | discSC | PED | MLC | DSC | Simple | Distinctive | Disaggregating |
|---|---|---|---|---|---|---|---|---|
| 1 | 53.6 | 52.2 | 54.1 | 56.4 | 46.0 | 60.2 | 60.7 | 62.0 |
| 2 | 57.8 | 60.4 | 64.3 | 60.9 | 49.2 | 70.0 | 70.0 | 72.0 |
| 3 | 41.3 | 40.0 | 40.4 | 30.2 | 31.7 | 42.1 | 42.9 | 46.5 |
| 4 | 58.0 | 56.3 | 68.7 | 60.3 | 50.9 | 75.3 | 76.2 | 76.8 |
| 6 | 62.7 | 54.1 | 54.9 | 50.9 | 54.5 | 60.4 | 61.1 | 62.7 |
| **Aggregate** | **54.7** | **52.6** | **56.5** | **51.7** | **46.5** | **61.6** | **62.2** | **64.0** |

TABLE II
TESTING MODE DISAGGREGATION ACCURACY

| **House** (trained on) | FHMM | discSC | PED | MLC | DSC | Simple | Distinctive | Disaggregating |
|---|---|---|---|---|---|---|---|---|
| 1 | 46.6 | 46.0 | 44.2 | 43.8 | 50.2 | 55.0 | 55.7 | 58.0 |
| 2 | 50.8 | 49.2 | 48.7 | 48.5 | 53.4 | 65.1 | 65.2 | 66.8 |
| 3 | 33.3 | 31.7 | 30.1 | 31.0 | 38.9 | 37.2 | 37.6 | 40.5 |
| 4 | 52.0 | 50.9 | 46.3 | 48.2 | 56.8 | 70.5 | 71.0 | 71.9 |
| 6 | 55.7 | 54.5 | 50.4 | 51.6 | 59.0 | 55.2 | 55.2 | 57.1 |
| **Aggregate** | **47.7** | **46.5** | **43.9** | **44.6** | **51.7** | **56.6** | **56.9** | **58.9** |

We see that between the three different proposals of ours, there is only light difference. The 'Simple' method yields good results. It is slightly improved with the 'Distinctive' penalty; the results improve further with the additional 'Disaggregating' penalty. The overall improvement we achieve over the existing techniques is 7.5%.

For the Training mode, we carried out statistical t-tests between the methods in order to verify if they are significantly different from each other. At 99% confidence interval, we found that our 'simple' and the 'distinctive' techniques were statistically similar but our final formulation – the 'disaggregating' technique was different (better). All of our proposed techniques were significantly better than the state-of-the-art.

### *B. Pecan Street Dataset*

We conduct this experiment on a subset of Dataport dataset available in NILMTK (non-intrusive load monitoring toolkit) format, which contains 1 minute circuit level and building level electricity data from 240 houses. The data set contains per minute readings from 18 different devices: air conditioner, kitchen appliances, electric vehicle, and electric hot tub heater, electric water heating appliance, dish washer, spin dryer, freezer, furnace, microwave, oven, electric pool heater, refrigerator, sockets, electric stove, waste disposal unit, security alarm and washer dryer. In the usual protocol about 80% of the homes are assigned as the training set and the remaining 20% of the homes as the test set. However in this work, we make the evaluation more challenging. We use 10% to 50% of the houses for training and the remaining for testing. The splitting into training and testing sets is done randomly and 100 such splits have been used in the experiments. What we report are the average of the 100 splits.

To prepare training and testing data, aggregated and sub-metered data are averaged over a time period of 10 minutes. This is the usual protocol to carry out experiments on the Pecan street dataset. Each training sample contains power consumed by a particular device in one day while each testing sample contains total power consumed in one day in particular house.

For our proposed techniques, the number of atoms for different appliances remain the same as before (i.e. three per appliance). The parametric values also remain the same as in REDD; we did not tune it any further. The configuration of the techniques compared against are also obtained from the non-intrusive load monitoring toolkit as before. The parameter settings from the state-of-the-art methods are from the corresponding papers.

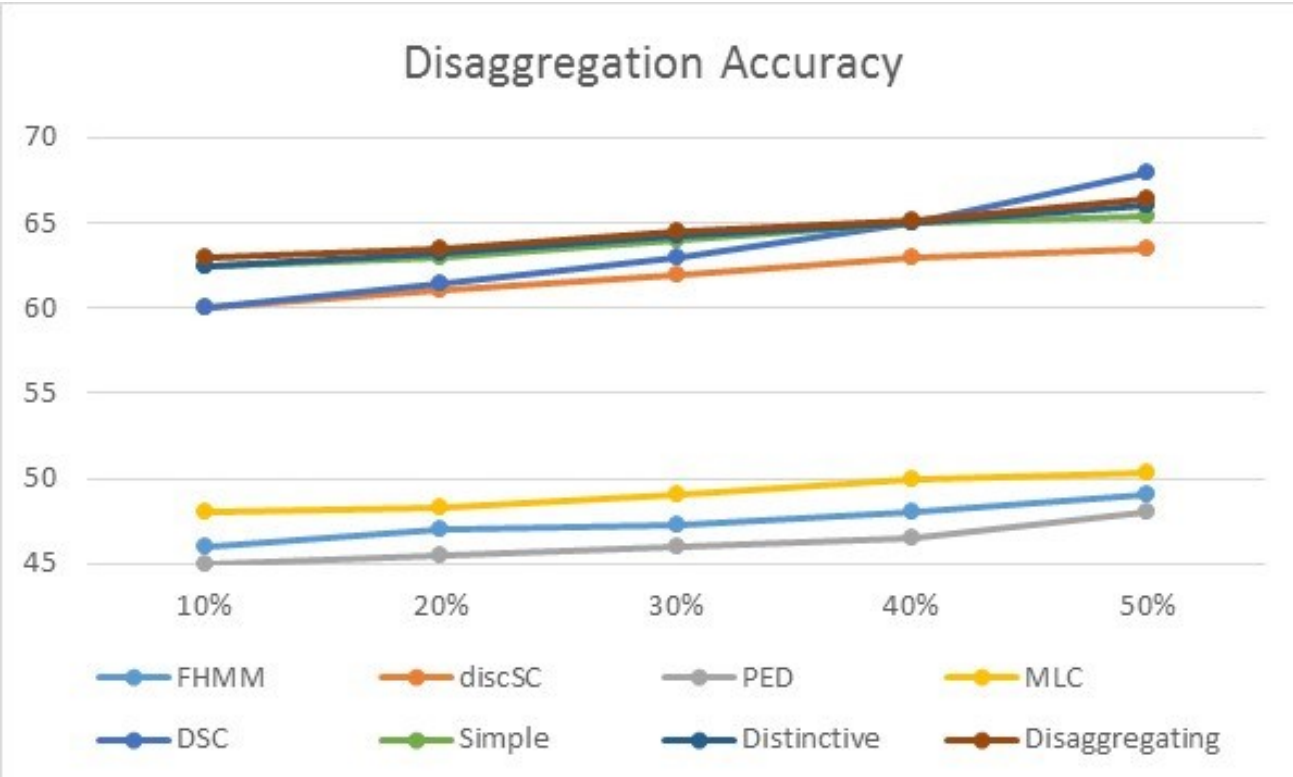

Fig. 1. Comparison of Disaggregation Accuracy

It is not possible to give the house-wise results like REDD. Therefore we show the results through two sets of graphs. The first graph (Fig. 1) shows the overall accuracy of each technique for a given training volume. We clearly see two distinct classes of techniques. The PED, MLC and FHMM are the bottom performing techniques; discSC, DSC and our proposed ones are better. Of these, discSC is the worst. DSC performs worse than ours when the volume of training data is low, but with increase the results continue to improve and eventually surpasses ours.

Notice that our proposed methods yield the same level of accuracy with only 10% training data as compared to state-of-the-art techniques utilizing 50% training data. This means that, given the scenario, we only need to instrument 10% of the homes as compared to 50% (required by existing methods); this is a drastic reduction in instrumentation and sensing cost. We refer to this result (five fold reduction in the need for sensing) in the introduction while giving the example.

The second set of graphs show the normalized error (a common metric) for common high power consuming appliances from different techniques. This is shown in Fig. 2. For this set of graphs we only show results for best performing methods – DSC and discSC; this is because the results from other techniques are so poor that the errors are larger by an order of magnitude making visual comparison meaningless.

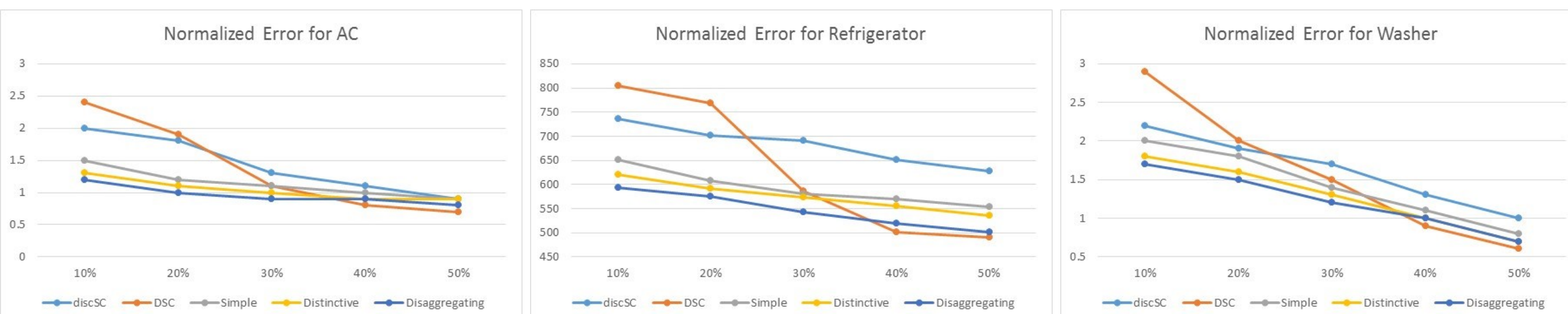

Fig. 2. Comparison of Normalized Error for (left to right) – AC, Refrigerator and Washer

The results show that, for smaller training volume our method performs the best. As the volume of training data increases deep sparse coding tends to perform better. This is expected. Deep learning overfits with small training data and hence perform poorly.

The conclusions drawn before hold in this set of plots as well. Our proposed method performs at par with state-of-the-art methods with far fewer training data; in practice this leads to far fewer instrumented homes, thus reducing the costs of sensors.

## V. Conclusion

This work proposes a new technique for energy disaggregation. It is based on analysis co-sparse coding. Results on benchmark databases show that the proposed technique performs better than others when the volume of training data is small. When the volume of training data is large, recently proposed method of deep sparse coding performs better.

In a practical scenario it would mean that our method will require far fewer number of instrumented houses, or far fewer days of instrumentation in each house for reaching the same level of accuracy as state-of-the-art techniques today. This means that using our method one can drastically reduce the sensing cost without losing on disaggregation accuracy.

## Acknowledgement

This work is partially supported by the Infosys Center for Artifical Intelligence at IIIT Delhi and by the DST IC-IMPACTS Indo-Canadian Grant.

## References

[1] G. Hart, "Nonintrusive appliance load monitoring", Proceedings of the IEEE, Vol. 80, pp. 1870-1891, 1992.

[2] J. Z. Kolter and T. S. Jaakkola, "Approximate Inference in Additive Factorial HMMs with Application to Energy Disaggregation", AISTATS, pp. 1472-1482, 2012.

[3] M. Dong; C. M. Paulo, W. Xu and C. Y. Chung, "Non-Intrusive Signature Extraction for Major Residential Loads", IEEE Transactions on Smart Grid, Vol. 4 (3), pp. 1421-1430, 2013.

[4] S. Makonin, F. Popowich, I. V. Bajic, B. Gill and L. Bartram, "Exploiting HMM Sparsity to Perform Online Real-Time Nonintrusive Load Monitoring", IEEE Transactions on Smart Grid, Vol. PP (99), pp. 1-11, 2015.

[5] J. Paris, J. S. Donnal and Steven B. Leeb, "NilmDB: The Non-Intrusive Load Monitor Database", IEEE Transactions on Smart Grid, Vol. 5 (5), pp. 2459 – 2467, 2014.

[6] Z. Kolter, S. Batra and A. Y. Ng. "Energy Disaggregation via Discriminative Sparse Coding", NIPS, pp. 1153-1161, 2011.

[7] E. Elhamifar and S. Sastry, "Energy Disaggregation via Learning 'Powerlets' and Sparse Coding", AAAI 2015.

[8] S. Singh and A. Majumdar, "Deep Sparse Coding for Non-Intrusive Load Monitoring", IEEE Transactions on Smart Grid , vol.PP, no.99, pp.1-1.

[9] F. Sultanem, "Using Appliance Signatures for Monitoring Residential Loads at Meter Panel Level", IEEE Transaction on Power Delivery, Vol. 6 (4), pp. 1380-1385, 1991.

[10] M.L. Marceau, R. Zmeureanu, "Nonintrusive load disaggregation computer program to estimate the energy consumption of major end uses for residential buildings", Energy Conversion & Management, Vol. 41, pp. 1389-1403, 2000.

[11] C. Laughman, K. D. Lee, R. Cox et al., "Advanced nonintrusive monitoring of electric loads," IEEE Power and Energy Magazine, pp. 56–563, April 2003.

[12] Y. Nakano, Non-Intrusive Electric Appliances Load Monitoring System Using Harmonic Pattern Recognition. Technical Report, 2004.

[13] S. Leeb, S. Shaw, J. Kirtley, "Transient Event Detection in Spectral Envelope Estimates for Nonintrusive Load Monitoring", IEEE Transactions on Power Delivery, Vol. 10 (3), pp. 1200-1210, 1995.

[14] J. G. Roos, I. E. Lane, E. C. Botha and G. P. Hancke, "Using neural networks for non-intrusive monitoring of industrial electrical loads," Instrumentation and Measurement Technology Conference, pp. 1115-1118, 1994.

[15] S. Patel, R. Robertson, J. A. Kinetz, M. S. Reynolds and G. D. Abowd, "At the Flick of a Switch: Detecting and Classifying Unique Electrical Events on the Residential Power Line", UbiComp, pp. 271–288, 2007.

[16] R. Cox, S. Leeb, S. Shaw, L. Norford, "Transient Event Detection for Nonintrusive Load Monitoring and Demand Side Management Using Voltage Distortion", IEEE Applied Power Electronics Conference and Exposition, pp. 7-10, 2006.

[17] T. D. Huang; Wen-Sheng Wang; Kuo-Lung Lian, "A New Power Signature for Nonintrusive Appliance Load Monitoring", IEEE Transactions on Smart Grid, Vol. 6 (4), pp. 1994 - 1995, 2015.

[18] T. Hassan, F. Javed and N. Arshad, "An Empirical Investigation of V-I Trajectory Based Load Signatures for Non-Intrusive Load Monitoring", IEEE Transactions on Smart Grid, Vol. 5 (2), pp. 870 - 878, 2014.

[19] M. Gulati, S. S. Ram, A. Majumdar and A. Singh, "Single Point Conducted EMI Sensor With Intelligent Inference for Detecting IT Appliances", IEEE Transactions on Smart Grid, 10.1109/TSG.2016.2639295.

[20] K. Basu, V. Debusschere, S. Bacha, U. Maulik and S. Bondyopadhyay, "Nonintrusive Load Monitoring: A Temporal Multilabel Classification Approach," IEEE Transactions on Industrial Informatics, Vol. 11 (1), pp. 262-270, 2015.

[21] D. Li and S. Dick, "Whole-house Non-Intrusive Appliance Load Monitoring via multi-label classification," 2016 International Joint Conference on Neural Networks (IJCNN), Vancouver, BC, 2016, pp. 2749-2755.

[22] S. M. Tabatabaei, S. Dick and W. Xu, "Toward Non-Intrusive Load Monitoring via Multi-Label Classification," in IEEE Transactions on Smart Grid, vol. 8, no. 1, pp. 26-40, Jan. 2017.

[23] B. Zhao, L. Stankovic and V. Stankovic, "On a Training-Less Solution for Non-Intrusive Appliance Load Monitoring Using Graph Signal Processing," in IEEE Access, vol. 4, no. , pp. 1784-1799, 2016.

[24] N Batra, H Wang, A Singh, K Whitehouse, "Matrix Factorisation for Scalable Energy Breakdown", AAAI 2017.

[25] N Batra, A Singh, K Whitehouse, "Gemello: Creating a detailed energy breakdown from just the monthly electricity bill", KDD 2016.

[26] J. Kelly, "Neural NILM: Deep Neural Networks Applied to Energy Disaggregation", ACM Buildsys, 2015.

[27] T. T. H. Le, J. Kim and H. Kim, "Classification performance using gated recurrent unit recurrent neural network on energy disaggregation," 2016 International Conference on Machine Learning and Cybernetics (ICMLC), Jeju, 2016, pp. 105-110.

[28] H. H. Chang, L. S. Lin, N. Chen and W. J. Lee, "Particle-Swarm-Optimization-Based Nonintrusive Demand Monitoring and Load Identification in Smart Meters," in IEEE Transactions on Industry Applications, vol. 49, no. 5, pp. 2229-2236, Sept.-Oct. 2013.

[29] Y. H. Lin and M. S. Tsai, "Non-Intrusive Load Monitoring by Novel Neuro-Fuzzy Classification Considering Uncertainties," in IEEE Transactions on Smart Grid, vol. 5, no. 5, pp. 2376-2384, Sept. 2014.

[30] A. Faustine, N. H. Mvungi, S. Kaijage and K. Michael, "A Survey on Non-Intrusive Load Monitoring Methodies and Techniques for Energy Disaggregation Problem", arXiv:1703.00785, 2017.

[31] R. Rubinstein, T. Peleg and M. Elad, "Analysis K-SVD: A Dictionary-Learning Algorithm for the Analysis Sparse Model," in IEEE Transactions on Signal Processing, vol. 61, no. 3, pp. 661-677, Feb.1, 2013.

[32] M. Aharon, M. Elad and A. Bruckstein, "K-SVD: An Algorithm for Designing Overcomplete Dictionaries for Sparse Representation", IEEE Transactions on Signal Processing, Vol. 54(11), pp. 4311-4322, 2006.

[33] W. Ruangsang and S. Aramvith, "Super-Resolution for HD to 4K using Analysis K-SVD dictionary and Adaptive Elastic-Net," 2015 IEEE International Conference on Digital Signal Processing (DSP), Singapore, 2015, pp. 1076-1080.

[34] A. Majumdar, "Improving Synthesis and Analysis Prior Blind Compressed Sensing with Low-rank Constraints for Dynamic MRI Reconstruction", Magnetic Resonance Imaging, Vol. 33(1), pp. 174-179, 2015.

[35] S. Boyd, N. Parikh, E. Chu, B. Peleato and J. Eckstein, "Distributed optimization and statistical learning via the alternating direction method of multipliers", Foundations and Trends® in Machine Learning, Vol. 3 (1), pp. 1-122, 2011.

[36] S. Boyd, (2011, December). Alternating direction method of multipliers. In Talk at NIPS Workshop on Optimization and Machine Learning.

[37] I. W. Selesnick and M. A. T. Figueiredo, "Signal restoration with overcomplete wavelet transforms: comparison of analysis and synthesis priors", Proceedings of SPIE, Vol. 7446 (Wavelets XIII), 2009.

[38] F. Li, J. F. P. J. Abascal, M. Desco and M. Soleimani, "Total Variation Regularization With Split Bregman-Based Method in Magnetic Induction Tomography Using Experimental Data," in IEEE Sensors Journal, vol. 17, no. 4, pp. 976-985, Feb.15, 15 2017.

[39] C. Bao, Y. Quan and H. Ji, "A Convergent Incoherent Dictionary Learning Algorithm for Sparse Coding", ECCV, 2014.

[40] J. Wang, J. F. Cai, Y. Shi and B. Yin, "Incoherent dictionary learning for sparse representation based image denoising," 2014 IEEE International Conference on Image Processing (ICIP), Paris, 2014, pp. 4582-4586.

[41] Tong Lin, Shi Liu, and Hongbin Zha, "Incoherent dictionary learning for sparse representation", IEEE ICPR, pp. 1237-1240, 2012.

[42] J. Z. Kolter and M. J. Johnson, "REDD: A public data set for energy disaggregation research", Proceedings of the SustKDD workshop on Data Mining Applications in Sustainability.

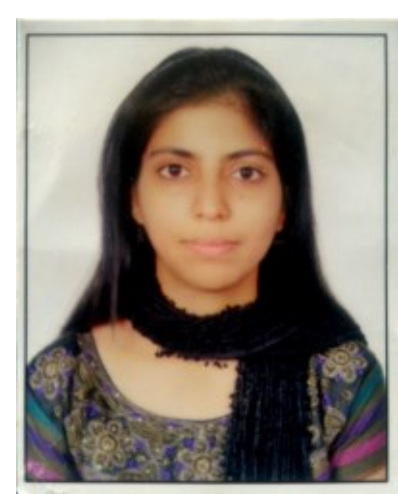

Shikha Singh did her B.Tech and M.Tech in Electronics and Communication from Guru Gobind Singh Indraprastha University. Currently, She is pursuing PhD at Indraprastha Institute of Information Technology. Her research includes Non Intrusive Load Monitoring and Machine Learning.

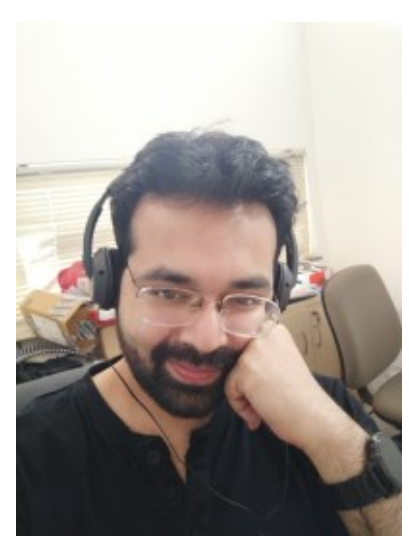

Angshul Majumdar got his Bachelor's degree from Bengal Engineering College, Shibpur in 2005. He did his Master's and PhD from the University of British Columbia in 2009 and 2012 respectively. His research interest is in optimizations, with applications in signal processing and machine learning. Currently his research interest lies in deep learning.

He is the founding chair of IEEE SPS Delhi Chapter and serving as the chair for the IEEE SPS chapter's committee from 2016-18.